\documentclass[conference]{IEEEtran}
\IEEEoverridecommandlockouts

\usepackage{cite}
\usepackage{amsmath,amssymb,amsfonts}
\usepackage{algorithmic}
\usepackage{graphicx}
\usepackage{textcomp}
\usepackage{amsmath}
\usepackage{adjustbox}
 
\def\BibTeX{{\rm B\kern-.05em{\sc i\kern-.025em b}\kern-.08em
    T\kern-.1667em\lower.7ex\hbox{E}\kern-.125emX}}
\usepackage[firstpage=true]{background}

\begin{document}
\SetBgContents{\parbox{\textwidth}{\footnotesize © 2025 IEEE. Personal use of this material is permitted. Permission from IEEE must be obtained for all other uses, in any current or future media, including reprinting/republishing this material for advertising or promotional purposes, creating new collective works, for resale or redistribution to servers or lists, or reuse of any copyrighted component of this work in other works.}}
\SetBgScale{1}
\SetBgAngle{0}
\SetBgPosition{current page.north}
\SetBgVshift{-1cm}
\SetBgColor{black}
\SetBgOpacity{1}

\title{Improved Motion Plane Adaptive 360-Degree Video Compression Using Affine Motion Models\\
}

\author{\IEEEauthorblockN{Marina Ritthaler, Andy Regensky, and André Kaup}
\IEEEauthorblockA{\textit{Multimedia Communications and Signal Processing} \\
\textit{Friedrich-Alexander-Universität Erlangen-Nürnberg}\\
Erlangen, Germany \\
marina.ritthaler@fau.de, andy.regensky@fau.de, and andre.kaup@fau.de}
\thanks{This work was supported by the Deutsche Forschungsgemeinschaft (DFG,
German Research Foundation) under project number 418866191.}
}
\maketitle

\begin{abstract}
Efficient compression of 360-degree video content requires the application of advanced motion models for interframe prediction. The Motion Plane Adaptive (MPA) motion model projects the frames on multiple perspective planes in the 3D space. It improves the motion compensation by estimating the motion on those planes with a translational diamond search. In this work, we enhance this motion model with an affine parameterization and motion estimation method. Thereby, we find a feasible trade-off between the quality of the reconstructed frames and the computational cost. The affine motion estimation is hereby done with the inverse compositional Lucas-Kanade algorithm. With the proposed method, it is possible to improve the motion compensation significantly, so that the motion compensated frame has a Weighted-to-Spherically-uniform Peak Signal-to-Noise Ratio (WS-PSNR) which is about 1.6 dB higher than with the conventional MPA. In a basic video codec, the improved inter prediction can lead to Bj\o ntegaard Delta (BD) rate savings between 9 \% and 35 \% depending on the block size (BS) and number of motion parameters.
\end{abstract}


\section{Introduction}
\begin{figure}[t]
    \centering
    \includegraphics[width=0.38\textwidth]{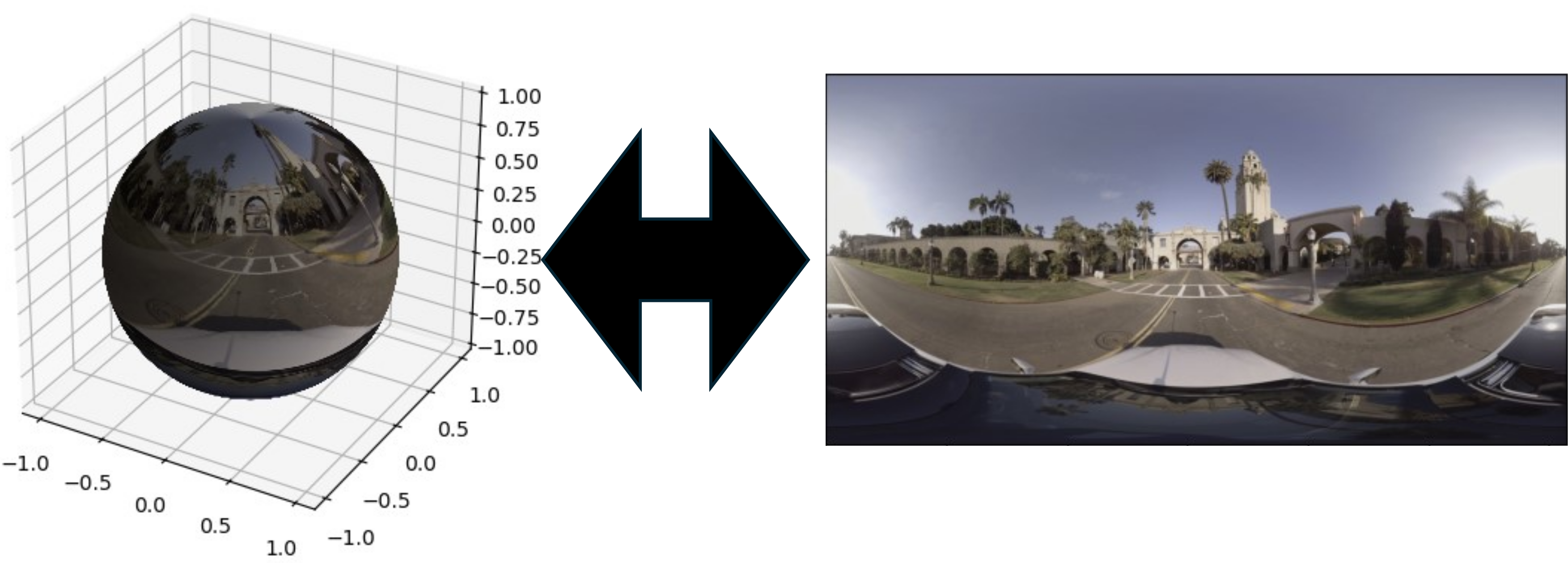}
    \caption{Spherical and equirectangular 360° video frame projection}
    \label{fig:projection}
\end{figure}

\noindent Latest technological advancements and a demand for an enriched user experience have led to
an expansion of the production and use of 360-degree videos. While this format is mostly
employed in a gaming and entertainment setting, it’s increasingly applied to education,
immersive telepresence, infotainment, and many more areas \cite{360sur}.
In order to use block-based video coding standards like High Efficiency Video Coding
(HEVC) \cite{HEVC} and Versatile Video Coding (VVC) \cite{VVC}, the omnidirectional video frames are usually projected to 2D formats before they are compressed. These projections lead to
geometry distortions and for some mappings to discontinuous face boundaries. As a consequence, the motion model of the blocks becomes irregular and the classic motion compensation does not perform well for 360-degree videos \cite{LiLi}. This performance decrease leads to a worse compression of those videos. Fig. \ref{fig:projection} shows an example image in a spherical and an equirectangular projection with its distortions.\\
\indent In \cite{model1,model2,model3,model4,model5,model6,model7,model8,regensky}, several strategies have been developed to improve the motion compensation for these frames. \cite{multi} evaluates multiple models and concludes that the highest Bj\o ntegaard Delta (BD) \cite{BD} rate savings are achieved with the Motion Plane Adaptive (MPA) motion modeling method from \cite{regensky}, where motion is modeled relative to a viewing position. So far, MPA is only able to represent translational motion. In this work, we enhance the approach with an affine parameterization and an affine motion estimation on the basis of the Lucas-Kanade (LK) algorithm. Despite significant quality improvements of affine MPA using the original LK motion estimation, it is infeasible in terms of complexity. In order to reach a significantly improved quality-complexity trade-off, we propose to utilize the inverse compositional LK algorithm and refine its adaptation in order to further decrease the computational complexity.

\section{Related Work}
\noindent Several methods have been proposed to improve the interframe prediction of 360-degree videos. The authors of \cite{model2} perform a translational block matching algorithm on a different perspective plane for each block. In \cite{model3}, Vishwanath et al. propose to estimate the motion of the block on the sphere by rotating the block position. \cite{model4} introduces a motion model, which calculates the relative movements of different points inside a block on the 2D projection, assuming an uniform motion on the sphere.\\
\indent The MPA model from \cite{regensky} utilizes multiple motion planes in the 3D space to perform the estimation and compensation. It assumes that the frames are in a 2D projection. In the first step, it projects the images onto the unit sphere. The pixel coordinates on the sphere are then rotated according to a rotation matrix. From the rotated sphere, a generalized perspective projection is obtained and used as motion plane. A translational motion vector is added to the pixel coordinates on this plane. The shifted coordinates are projected back with the inverse of the projection that was performed beforehand. The authors propose to use three different rotation matrices, which yield the planes for front/back, left/right and top/bottom.\\
\indent For regular video content, affine motion estimation is already used to improve the compression \cite{aff1, aff2}. In contrast, for motion models of 360-degree videos, there exists only translational estimation. Therefore, we aim at extending MPA with an affine model and evaluating the results.

\section{Affine Motion Plane Adaptive Motion Estimation}
\begin{figure}[t]
    \centering
    \includegraphics[width=0.25\textwidth]{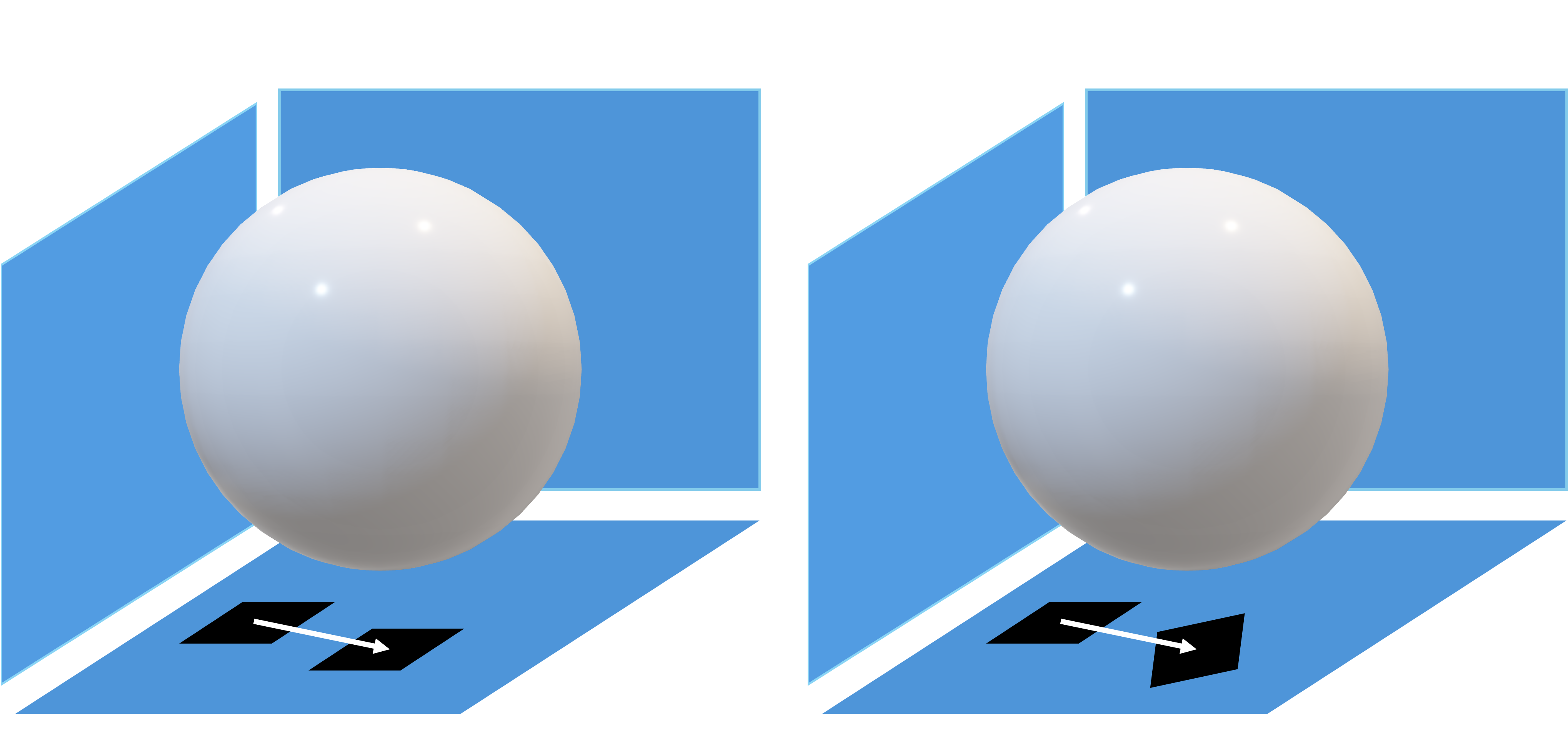}
    \caption{Translational and affine MPA visualization}
    \label{fig:MPAVis}
\end{figure}

\noindent In this work, we enhance the algorithm with a motion model that allows an affine motion of the blocks on the motion planes. This is shown in Fig. \ref{fig:MPAVis}. The left side visualizes the original MPA and the right side pictures our extension. The affine version of MPA is done with a six and a four parameter model. Using only four parameters still enables any combination of zooming, rotation and translation. It allows less motion than the standard affine transform, but has also fewer motion parameters that need to be estimated and transmitted.
\\
\indent The algorithm introduced in this work starts by reading in a reference frame and the frame that has to be transmitted, which will be called the template frame. The template frame is split into blocks. The conventional translational MPA parameters serve as a starting point for the affine motion estimation. The motion model used in this method is
\begin{equation}
    m_{mpa}(\mathbf{x}_{e}, \mathbf{t}, \mathbf{R}) = \zeta_{\mathbf{R}}^{-1}(\zeta_{\mathbf{R}}(\mathbf{x}_{e})+\mathbf{t}),
    \label{eq:MPA_projection}
\end{equation}
with the pixel coordinates of the reference frame $\mathbf{x}_{e}$ and the translational motion vector $\mathbf{t}$. Using $\mathbf{R}$ to describe the desired motion plane, $\zeta_{\mathbf{R}}$ and $\zeta_{\mathbf{R}}^{-1}$ define the full transform from the 2D frame to the plane and its inverse.\\
\indent Following this, the affine estimation is performed according to the inverse compositional algorithm (IC). The update step size of the algorithm is increased in order to escape more local minima. Because of that, an even better result is achieved. For a reduction of the computation time, the motion estimation is only performed on one of the perspective planes, while the original MPA method uses all three planes and chooses the one, on which it reaches the lowest error. 
\subsection{Application of the Inverse Compositional Lucas-Kanade for Motion Estimation}
\label{subsec:ICL}
\noindent The MPA method from \cite{regensky} performs a block matching using a diamond search algorithm. This works very well with translational motion, because it has only two degrees of freedom (DoF). With increased DoF, this method becomes infeasible, as there would be to many possible motions to check.\\ 
\indent One approach to find the motion between frames is via optical flow. One of the most widely used techniques for optical flow estimation is the LK algorithm introduced in \cite{LK}. It aims at aligning a template $\mathbf{T}(\mathbf{x})$ and an image $\mathbf{I}(\mathbf{x})$ with the pixel coordinates $\mathbf{x} = (x, y)^{T}$. For block motion estimation, the template corresponds to the pixels in the current block of the current frame and the image is taken from the reference frame. The algorithm minimizes the sum of squared differences (SSD) between the template and a warped image. The minimization is done with respect to the motion parameters $\mathbf{p}$ of the parameterized set of allowed warps W. The warps are mapping functions for the pixels. They can be arbitrarily complex with an arbitrarily large number of parameters. The optimization is performed iteratively. Therefore, only small parameter changes are computed in each iteration, and the expression
\begin{equation}
	\sum_{x}[\mathbf{I}(W(\mathbf{x}; \mathbf{p} + \Delta \mathbf{p})) - \mathbf{T}(\mathbf{x})]^{2}
\end{equation}
is linearized and minimized with respect to $\Delta \mathbf{p}$ using least squares. At the end of the iteration, an update of the parameters is performed by adding $\Delta \mathbf{p}$ to the current parameters.
\\
\indent A version of the LK, which decreases its complexity is the IC \cite{LK_IC}. It reduces the computational cost of the original LK from $\mathcal{O}(n^{2}UV + n^{3})$ to $\mathcal{O}(nUV + n^{3})$. $n$ is hereby the number of parameters, $U$ the width, and $V$ the height of the template. While the classic method requires to evaluate a Jacobian and a Hessian matrix in every iteration, the IC does it only once in the beginning of the algorithm \cite{LK_20}.
This is achieved by making two changes. Firstly, it inverts the roles of the image and template. Secondly, instead of an additive update to the parameters, it uses an incremental warp $W(\mathbf{x};\Delta \mathbf{p})$ and performs the update with
\begin{equation}
	W(\mathbf{x}; \mathbf{p}) \leftarrow W(\mathbf{x}; \mathbf{p}) \circ W(\mathbf{x}; \Delta \mathbf{p})^{-1}.
\end{equation}
In the update, $W(\mathbf{x}; \mathbf{p}) \circ  W(\mathbf{x};\Delta \mathbf{p})^{-1}$ stands for $W(W(\mathbf{x};\Delta \mathbf{p})^{-1}; \mathbf{p})$. 
The equation that needs to be minimized is now given by
\begin{equation}
	\sum_{\mathbf{x}}[\mathbf{T}(W(\mathbf{x}; \Delta \mathbf{p})) - \mathbf{I}(W(\mathbf{x}; \mathbf{p}))]^{2}.
\end{equation}
\noindent When applying the original LK in the MPA motion estimation, the computational cost becomes infeasibly high. Therefore, only a use of the IC is applicable. In this method, which is denoted by MPA\textsubscript{IC}, the parameter optimization is done for a warp, which includes the projections, as well as the affine transform.\\
\indent An affine transform is performed by multiplying the coordinates with a transform matrix $\mathbf{A}(\mathbf{p})$ with either six or four DoF,
\begin{equation}
    \mathbf{x}_{new} = \mathbf{A}(\mathbf{p}) \mathbf{x}.
\end{equation}
The corresponding transform matrix for six parameters is
\begin{equation}
    \mathbf{A}_{6}(\mathbf{p_{6}}) = \begin{bmatrix}
    a+1 & b & e\\
    c & d+1 & f\\
    0 & 0 & 1
    \end{bmatrix},
\end{equation}
and the one for four parameters is
\begin{equation}
    \mathbf{A}_{4}(\mathbf{p_{4}}) = \begin{bmatrix}
    a+1 & b & e\\
    -b & a+1 & f\\
    0 & 0 & 1
    \end{bmatrix}.
\end{equation}
For the warp, we obtain
\begin{equation}
    \label{eq:MPAwarp}
    W(\mathbf{x};\mathbf{p})=\zeta_{\mathbf{R}}^{-1}(\mathbf{A}(\mathbf{p})\zeta_{\mathbf{R}}(\mathbf{x})).
\end{equation}
The IC performs its update as
\begin{equation}
    W(\mathbf{x};\mathbf{p}) \leftarrow W(W(\mathbf{x};\Delta \mathbf{p})^{-1};\mathbf{p}).
    \label{eq:inv_comp_warp}
\end{equation}
Hence, the warp must be inverted. Since the warp contains the projections and is therefore rather complicated, we reduce the complexity of its inversion in the following.\\
\indent An inverse warp satisfies $\mathbf{x}=W(W(\mathbf{x};\mathbf{p})^{-1};\mathbf{p})$. When performing the inverse of a projection function upon that projection on some coordinates, the coordinates stay the same. Therefore, with the identity matrix $\mathbf{I}$, we have
\begin{equation}
        \mathbf{x} = \zeta_{\mathbf{R}}^{-1}(\mathbf{I}\zeta_{\mathbf{R}}(\mathbf{x})).
    \label{eq:inverse_ex1}
\end{equation}
From there, it is possible to replace the identity matrix with $\mathbf{A}(\mathbf{p})\mathbf{A}^{-1}(\mathbf{p})$. In a second step, one can apply the projection function and its inverse after performing the inverse affine transform to get
\begin{equation}
    \begin{split}
        \mathbf{x} &= \zeta_{\mathbf{R}}^{-1}(\mathbf{A}(\mathbf{p})\mathbf{A}^{-1}(\mathbf{p})\zeta_{\mathbf{R}}(\mathbf{x}))\\
        &= \zeta_{\mathbf{R}}^{-1}(\mathbf{A}(\mathbf{p})\zeta_{\mathbf{R}}(\zeta_{\mathbf{R}}^{-1}(\mathbf{A}^{-1}(\mathbf{p})\zeta_{\mathbf{R}}(\mathbf{x}))))\\
        &= W(W(\mathbf{x};\mathbf{p})^{-1};\mathbf{p}).
    \end{split}
    \label{eq:inverse_ex2}
\end{equation}
The inverse warp can therefore be simplified to
\begin{equation}
    \begin{split}
    W(\mathbf{x};\Delta \mathbf{p})^{-1} &= (\zeta_{\mathbf{R}}^{-1}(\mathbf{A}(\Delta \mathbf{p})\zeta_{\mathbf{R}}(\mathbf{x})))^{-1}\\
    &= \zeta_{\mathbf{R}}^{-1}(\mathbf{A}^{-1}(\Delta \mathbf{p})\zeta_{\mathbf{R}}(\mathbf{x})).
    \end{split}
    \label{eq:inv_comp_inverse_warp}
\end{equation}
Since the inverse of the warp equals the warp with an inverted affine transform, the update of the warp can be computed with
\begin{equation}
    \begin{split}
    &W(\mathbf{x};\mathbf{p}) \leftarrow \zeta_{\mathbf{R}}^{-1}(\mathbf{A}(\mathbf{p})\zeta_{\mathbf{R}}(\zeta_{\mathbf{R}}^{-1}(\mathbf{A}^{-1}(\Delta \mathbf{p})\zeta_{\mathbf{R}}(\mathbf{x}))))\\
    &W(\mathbf{x};\mathbf{p}) \leftarrow \zeta_{\mathbf{R}}^{-1}(\mathbf{A}(\mathbf{p})\mathbf{A}^{-1}(\Delta \mathbf{p})\zeta_{\mathbf{R}}(\mathbf{x})).
    \end{split}
    \label{eq:inv_comp_warp_full}
\end{equation}
\\
Updating the warp therefore only requires the calculation of a new affine matrix
\begin{equation}
    \mathbf{A^{*}}(\mathbf{p}) = \mathbf{A}(\mathbf{p})\mathbf{A}^{-1}(\Delta \mathbf{p})
\end{equation}
\subsection{Motion Compensation and Evaluation}
\noindent With the known motion parameters, an estimate of the current frame can be computed. For the motion compensated frame, the warp from equation \eqref{eq:MPAwarp} is applied with the calculated parameters. For each block, a prediction image is interpolated with coordinates that are computed with the MPA model. Therefore, the block coordinates are transformed onto the perspective plane that belongs to the motion parameters of that block and the affine transform is applied. Subsequently, the coordinates are transformed back and used for the interpolation. With the predicted blocks $\mathbf{B}_{ij}$ and the reference frame $\mathbf{I_{ref}}$, we have\\
\begin{equation}
    \mathbf{B}_{ij} = \mathbf{I_{ref}}(W(\mathbf{x}_{ij}; \mathbf{p}_{ij})).
\end{equation}
When this is performed for every block position for the predicted image, the full frame gets motion compensated
\begin{equation}
    \mathbf{I_{pred}} = 
    \begin{bmatrix} 
    \mathbf{B}_{11} & \dots & \mathbf{B}_{1J} \\
    \vdots & \ddots & \vdots \\
    \mathbf{B}_{I1} & \dots & \mathbf{B}_{IJ} 
    \end{bmatrix}.
\end{equation}
Then, the difference between the approximated and the actual frame, and the quality of the estimation method, can be evaluated.
\section{Performance Evaluation}

\begin{table*}[t]
\caption{psnr [$\scriptstyle\mathrm{dB}$], ws-psnr [$\scriptstyle\mathrm{dB}$] and computation time of the motion compensated frames for each method. the mpa\textsubscript{translational} results serve as anchor for the relative values of the other methods.}
\label{Tab: comp}
\begin{center}
\begin{adjustbox}{max width=\linewidth}
\begin{tabular}{|c|c||c|c|c||c|c|c||c|c|c||c|c|c|}
\hline
\textbf{ }&\textbf{ }&\multicolumn{3}{|c||}{\textbf{\footnotesize TMC} \cite{TMC}}&\multicolumn{3}{|c||}{\textbf{\footnotesize MPA\textsubscript{translational}} \cite{regensky}}&\multicolumn{3}{|c||}{\textbf{\footnotesize MPA\textsubscript{IC6P}}} &\multicolumn{3}{|c|}{\textbf{\footnotesize MPA\textsubscript{IC4P}}} \\
\scriptsize BS& \scriptsize video & \scriptsize PSNR& \scriptsize WS-PSNR& \scriptsize Time&\scriptsize PSNR& \scriptsize WS-PSNR& \scriptsize Time& \scriptsize PSNR& \scriptsize WS-PSNR&\scriptsize Time&\scriptsize PSNR& \scriptsize WS-PSNR& \scriptsize Time\\
\hline

&\scriptsize Balboa&\footnotesize -0.93 &\footnotesize -0.83 &\footnotesize -86 \%&\footnotesize 42.52&\footnotesize 41.66&\footnotesize 100 \%&\footnotesize +0.83 &\footnotesize +0.86 &\footnotesize +50 \%&\footnotesize +0.48 &\footnotesize +0.47 &\footnotesize +54 \%\\

&\scriptsize BranCastle2& \footnotesize -1.65&\footnotesize -0.97& \footnotesize -87 \%& \footnotesize 33.34& \footnotesize 33.72& \footnotesize 100 \%& \footnotesize +1.28& \footnotesize +1.30& \footnotesize +42 \%& \footnotesize +0.98& \footnotesize +0.88& \footnotesize +45 \%\\

&\scriptsize Broadway& \footnotesize -0.84& \footnotesize -0.82& \footnotesize -86 \%& \footnotesize 39.66& \footnotesize 38.47& \footnotesize 100 \%& \footnotesize +1.05& \footnotesize +1.07& \footnotesize +53 \%& \footnotesize +0.56& \footnotesize +0.56& \footnotesize +57 \%\\

\footnotesize 16&\scriptsize ChairliftRide& \footnotesize -2.19& \footnotesize -1.37& \footnotesize -86 \%& \footnotesize 44.87& \footnotesize 44.68& \footnotesize 100 \%& \footnotesize +0.61& \footnotesize +0.83& \footnotesize +51 \%& \footnotesize +0.37& \footnotesize +0.47&\footnotesize +55 \%\\

&\scriptsize Landing2&\footnotesize -0.45&\footnotesize -0.29&\footnotesize -87 \%&\footnotesize 34.37&\footnotesize 33.51&\footnotesize 100 \%&\footnotesize +1.22&\footnotesize +1.23&\footnotesize +51 \%&\footnotesize +0.78&\footnotesize +0.78&\footnotesize +54 \%\\

&\scriptsize SkateboardInLot&\footnotesize -0.48&\footnotesize -0.47&\footnotesize  -87 \%&\footnotesize 36.69&\footnotesize 35.43&\footnotesize 100 \%&\footnotesize +1.39&\footnotesize +1.54&\footnotesize +41 \%&\footnotesize +0.66&\footnotesize +0.71&\footnotesize +44 \%\\

\cline{2-14}
&\scriptsize Average&\footnotesize -1.09 &\footnotesize -0.79&\footnotesize  -86 \%&\footnotesize 38.58 &\footnotesize 37.91 &\footnotesize 100 \%&\footnotesize +1.06 &\footnotesize +1.14 &\footnotesize +48 \%&\footnotesize +0.63 &\footnotesize +0.65 &\footnotesize +52 \%\\
\hline

&\scriptsize Balboa&\footnotesize -1.07 &\footnotesize -1.01 &\footnotesize -80 \%&\footnotesize 40.72&\footnotesize 39.86&\footnotesize 100 \%&\footnotesize +1.60 &\footnotesize +1.65 &\footnotesize +111 \%&\footnotesize +0.84 &\footnotesize +0.85 &\footnotesize +117 \%\\

&\scriptsize BranCastle2& \footnotesize -1.66 & \footnotesize -0.98 & \footnotesize -82 \%& \footnotesize 31.66 & \footnotesize 32.15 & \footnotesize 100 \%& \footnotesize +2.10 & \footnotesize +2.09 & \footnotesize +118 \%& \footnotesize +1.54 & \footnotesize +1.42 & \footnotesize +122 \%\\

&\scriptsize Broadway& \footnotesize -0.96& \footnotesize -0.99 & \footnotesize -80 \%& \footnotesize 37.76& \footnotesize 36.57& \footnotesize 100 \%& \footnotesize +1.39& \footnotesize +1.39& \footnotesize +115 \%& \footnotesize +0.55& \footnotesize +0.54& \footnotesize +120 \%\\

\footnotesize  32&\scriptsize ChairliftRide& \footnotesize -2.91& \footnotesize -1.78 & \footnotesize -80 \%& \footnotesize 43.92& \footnotesize 43.48 & \footnotesize 100 \%& \footnotesize +1.34 & \footnotesize +1.22 & \footnotesize +109 \%& \footnotesize +0.87 & \footnotesize +0.69 & \footnotesize +114 \%\\

&\scriptsize Landing2&\footnotesize -0.49 &\footnotesize -0.30 &\footnotesize -81 \%&\footnotesize 32.89&\footnotesize 32.05&\footnotesize 100 \%&\footnotesize +1.58 &\footnotesize +1.62 &\footnotesize +117 \%&\footnotesize +0.90 &\footnotesize +0.92 &\footnotesize +121 \%\\

&\scriptsize SkateboardInLot&\footnotesize -0.54 &\footnotesize -0.54 &\footnotesize  -82 \%&\footnotesize 34.89 &\footnotesize 33.52&\footnotesize 100 \%&\footnotesize +1.75 &\footnotesize +1.95 &\footnotesize +102 \%&\footnotesize +0.74 &\footnotesize +0.81 &\footnotesize +105 \%\\

\cline{2-14}
&\scriptsize Average&\footnotesize -1.27 &\footnotesize -0.93 &\footnotesize -81 \%&\footnotesize 36.97&\footnotesize 36.27 &\footnotesize 100 \%&\footnotesize +1.63 &\footnotesize +1.66 &\footnotesize +112 \%&\footnotesize +0.91 &\footnotesize +0.87 &\footnotesize +117 \%\\
\hline
\end{tabular}
\label{tab1}
\end{adjustbox}
\end{center}
\end{table*}
\subsection{Experimental Setup}
\noindent The methods are tested on six grayscale video sequences from the JVET common test conditions \cite{testcond} with frame rates of 30 and 60 frames per second. Their bit-depth is eight and ten bits. To decrease the time for testing, the frames are downscaled to $384 \times 768$ pixels. From each video, the results are averaged over 32 frames distributed evenly over the video. The template and the reference are always two consecutive frames. As quality metric for the motion estimation, the Peak Signal-to-Noise Ratio (PSNR) and the Weighted-to-Spherically-uniform Peak Signal-to-Noise Ratio (WS-PSNR) \cite{WSPSNR} are used. Additionally, the computation time, normalized to the computation time of the translational MPA method, is evaluated.\\
\indent The results of the methods are compared to the translational MPA method and a diamond search block matching method that is performed on the equirectangular projection of the frames, denoted by translational motion compensation (TMC). For the block matching, the search range is set to 96 pixels. The subpixel precision applied is $1/8$ of a pixel.
\subsection{Motion Compensation and Compression Improvements}
\begin{figure}[t]
    \centering
    \includegraphics[width=0.46\textwidth]{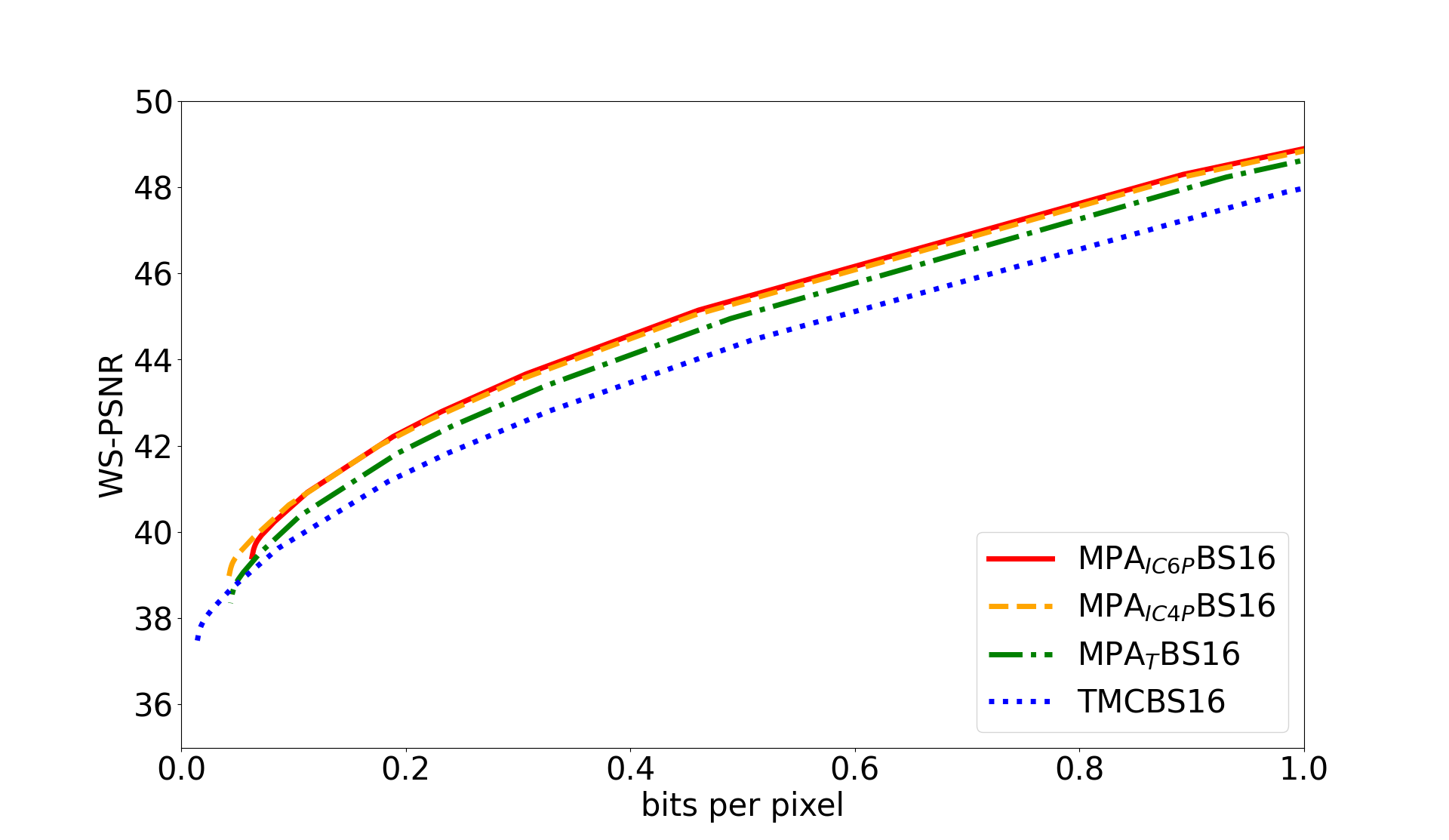}
    \caption{WS-PSNR over bits per pixel averaged over all videos for BS 16}
    \label{fig:rate16}
\end{figure}

\begin{figure}[t]
    \centering
    \includegraphics[width=0.46\textwidth]{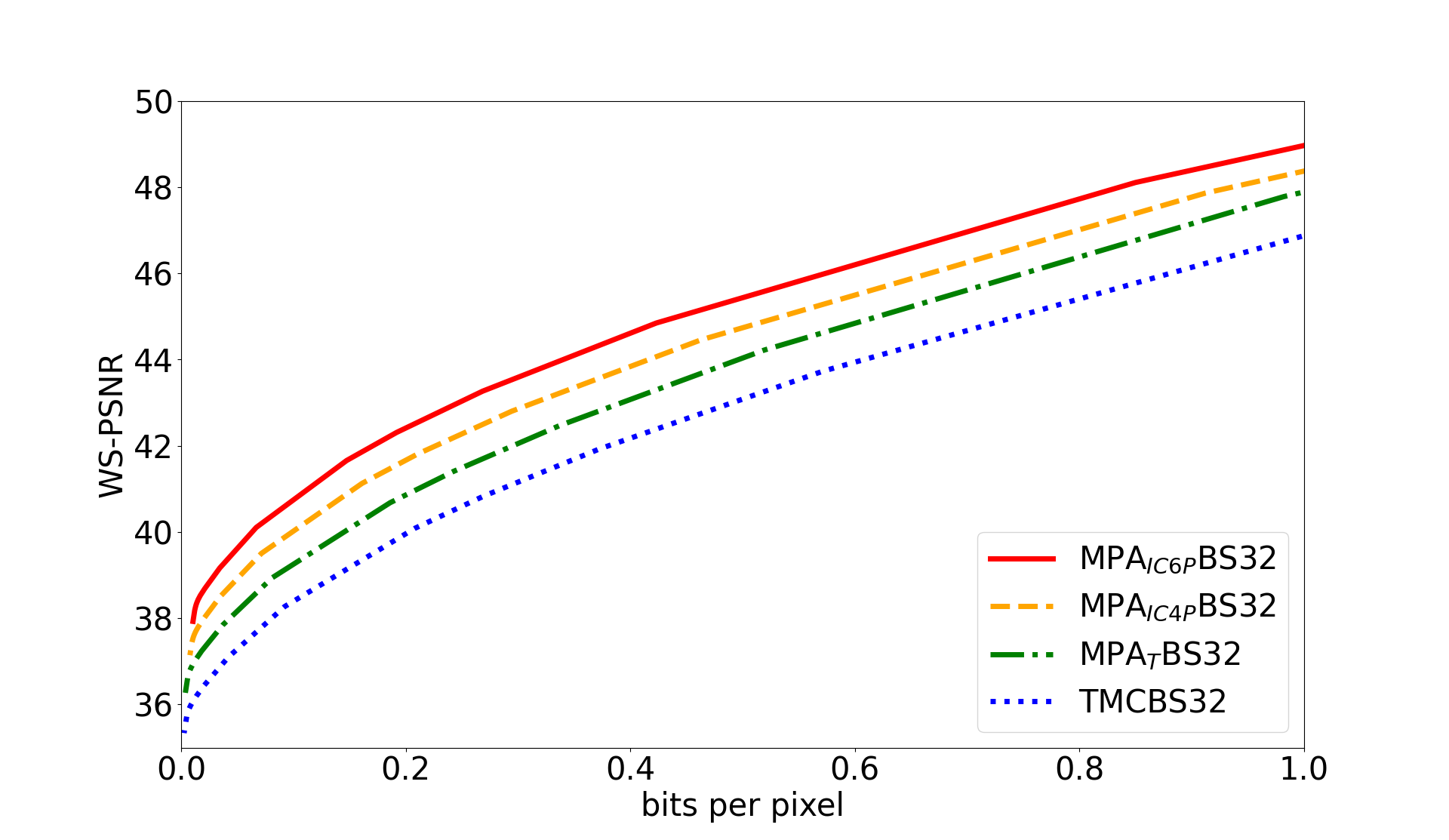}
    \caption{WS-PSNR over bits per pixel averaged over all videos for BS 32}
    \label{fig:rate32}
\end{figure}

\noindent Table \ref{Tab: comp} gives the averaged motion compensation results for the four videos for the four methods in dB. The results of the MPA\textsubscript{translational} are presented in their absolute value. For the other methods, the results are given relative to the MPA\textsubscript{translational}. For a block size (BS) of 16, the six parameter affine method increases the PSNR and WS-PSNR by about 1.1 dB compared to the translational MPA method. The four parameter model can still reach an increase of 0.6 dB. Hereby, the computation time increases by around 50 \%. For the bigger BS, the quality metrics even increase by more than 1.6 dB for the six, and around 0.9 dB for the four parameter model. In this case, the computation time doubles compared to the original MPA method.\\
\indent Similar to \cite{model8}, we implemented a basic video codec to estimate the bitrate savings for the same quality on a rate distortion curve. The codec gets the template frame, reference frame and the motion parameters as input. The error frame between the template and the motion compensated reference frame is calculated. Subsequently, this error frame is divided into blocks. A discrete cosine transform is performed on each block and the result is quantized with a quantization matrix that is mostly focused on the low frequencies. The quantized values are transformed to levels and the block is zigzag scanned and represented by runlevels. Lastly, the runlevels are Huffman encoded. To get an estimate on the bit length of the motion parameters, the bzip2 package \cite{bzip}, which performs a lossless compression, is employed. The encoder has a quantization parameter, which is multiplied with the quantization matrix. With different parameter values, different coding qualities are reached.\\
\indent Figures \ref{fig:rate16} and \ref{fig:rate32} show the WS-PSNR over the rate in bits per pixel for all methods for a BS of 16 and a BS of 32 averaged over 32 frames of all videos. With our codec, the curves only show the compression of interframe images and assume that the reference frame is already known at the decoder. For both BSs the error image bit savings outweigh the increase in bit stream length due to the higher number of motion parameters. Especially for higher BSs, the total and therefore also the increase in motion parameter bits becomes negligibly small. For a BS of 32, the six parameter model outperforms the four DoF model with BD rate savings of 35.2 \% compared to the original MPA, while the four parameter model achieves to reduce the BD rate by 21.0 \%. For BS 16, having less motion parameters gives the better result with rate savings of 9.5 \%, while the six DoF method reaches a reduction of 9.0 \% In general the affine method reaches a better improvement with higher BSs.

\section{Conclusion}
\noindent With the extension of the MPA by an affine parameterization and an efficient motion estimation, it is possible to achieve a significant improvement of quality of the motion compensated frame. Hereby, the enhancement of the WS-PSNR and PSNR reaches more than 1.6 dB for a BS of 32. With the improved motion estimation, a better rate distortion performance with BD rate savings of 9.0 to 35.2 \% can be achieved. As drawback, we have an increase in computational complexity for affine techniques. However, with our method, this rise can be limited to twice the time of the translational MPA method.\\ 
\indent The next step is to integrate the affine MPA method into a standardized video codec, such as VVC. This way, the exact rate reduction that can be achieved with affine motion in a rate distortion optimized setting can be found. It is expected to be even larger.

\end{document}